\documentclass[11pt]{article}

\usepackage[portrait, margin=1.25in]{geometry}
\usepackage{setspace}
\usepackage[utf8]{inputenc} % allow utf-8 input
\usepackage[T1]{fontenc}    % use 8-bit T1 fonts
\usepackage{hyperref}       % hyperlinks
\usepackage{url}            % simple URL typesetting
\usepackage{booktabs}       % professional-quality tables
\usepackage{amsfonts}       % blackboard math symbols
\usepackage{nicefrac}       % compact symbols for 1/2, etc.
\usepackage{microtype}      % microtypography

\usepackage{graphicx}
\usepackage[ruled,vlined]{algorithm2e}
\usepackage{indentfirst}
\usepackage{scalerel,stackengine}
\stackMath
\newcommand\reallywidehat[1]{%
\savestack{\tmpbox}{\stretchto{%
  \scaleto{%
    \scalerel*[\widthof{\ensuremath{#1}}]{\kern-.6pt\bigwedge\kern-.6pt}%
    {\rule[-\textheight/2]{1ex}{\textheight}}%WIDTH-LIMITED BIG WEDGE
  }{\textheight}% 
}{0.5ex}}%
\stackon[1pt]{#1}{\tmpbox}%
}
\usepackage{appendix}
\usepackage{multicol}
\usepackage{amsmath}

\DeclareMathOperator*{\argmin}{arg\,min}

\usepackage{caption}
\usepackage{subcaption}
\usepackage{multirow}
\usepackage{ragged2e}
\usepackage{ dsfont }
\usepackage{amsthm}
\usepackage{relsize}

\usepackage{authblk}

\title{\Large Machine Learning Portfolio Allocation}

\author[*]{Michael Pinelis}
\author[**]{David Ruppert}
\affil[*]{Department of Economics, Cornell University, \texttt{mdp93@cornell.edu}}
\affil[**]{Department of Statistics \& Data Science and School of Operations Research and Information Engineering, Cornell University, \texttt{dr24@cornell.edu}}

\date{March 2, 2020}

\begin{document}
\maketitle

\begin{abstract}
We find economically and statistically significant gains when using machine learning for portfolio allocation between the market index and risk-free asset. Optimal portfolio rules for time-varying expected returns and volatility are implemented with two Random Forest models. One model is employed in forecasting monthly excess returns with macroeconomic factors including payout yields. The second is used to estimate the prevailing volatility. Reward-risk timing with machine learning provides substantial improvements over the buy-and-hold in utility, risk-adjusted returns, and maximum drawdowns. This paper presents a unifying framework for machine learning applied to both return- and volatility-timing.

\end{abstract}

\newline \indent  market timing, reward-risk timing, volatility estimation, equity return predictability


%\begin{multicols}{2}
\newpage

\section{Introduction}

\noindent We use machine learning to find the optimal portfolio weights between the market index and the risk-free asset. The timing strategy is generated from the utility maximization principle and gives optimal portfolio weights estimated monthly with two Random Forest models. The market weight is proportional to the reward factor, which is a forecast of the excess market return\footnote{We refer to excess market return as the excess over the risk-free rate in this paper.}, and is inversely proportional to the risk factor, an estimate of prevailing squared volatility. This procedure is simultaneously return- and volatility-timing the market and can be called 'reward-risk timing'\footnote{This term is from Kirby and Ostdiek (2012), who propose weighting by individual price of risks in a multi-asset portfolio. Our paper focuses on the portfolio with the market index and risk-free asset. Another difference is Kirby and Ostdiek (2012) use several-year-long rolling window estimates of the conditional mean and volatility while we look at short-term windows for machine learning strategies.}. Our method found that a portfolio allocation strategy employing machine learning to reward-risk time the market gave significant improvements in investor utility and Sharpe ratios and earned a large alpha of 3.4\%. We motivate our analysis from the vantage point of a utility-maximizing investor, who adjusts the portfolio allocation according to the attractiveness of the risk-reward trade-off. %The results of this paper can be applied by industry practitioners, institutional investors, or the individual investor.

A number of papers have been written on predicting returns and volatilities with machine learning and large numbers of features. See as a review (Henrique et al., 2019). Machine learning methods have been shown to be suitable and advantageous for the difficult task of identifying the regimes in the markets (Gu et al., 2020). Gu et al.\ find a benefit of using machine learning for market timing with return forecasts of 26\% and 18\% increases in Sharpe ratios with neural networks and Random Forest, respectively, relative to that of the buy-hold. Yet none predict returns and volatilities with machine learning in combination. Our results document a 28\% increase in Sharpe ratios when using Random Forest for both returns and volatilities in combination. Taking advantage of the allowance for nonlinear predictor interactions in machine learning models gives better return and volatility forecasts based on market conditions. An approach with machine learning that considers both expected return- and volatility-timing leads to a profitable trading strategy, without an extensive set of predictors. This paper studies how the machine learning methods of Random Forest and Elastic Net can forecast the excess return with the major conditioning variables proposed so far in the literature and summarized by Goyal and Welch (2008) as well as an enhanced measure of the payout yield (see Boudoukh et al. (2007)). Then separate Random Forest and Elastic Net models are employed to predict next month's volatility with the similar set of variables. Comparing the performance of a standard linear model for reward-risk timing, we show that the machine learning models outperform by a significant margin.

Expected-return or reward-timing involves adjusting the portfolio allocation according to beliefs about future asset returns. This is akin to benchmark timing, the active management decision to vary the managed portfolio's beta with respect to the benchmark (Grinold and Kahn, 1999). Merton (1981) derived the economic value of return forecasts. Campbell and Thompson (2008) show that many predictive regressions beat the historical average return, once weak restrictions are imposed on the signs of coefficients and return forecast.

Volatility- or risk-timing is a newer idea. While there is a wide array of volatility-based portfolio allocation strategies, this paper's trading rule is derived from the utility maximization principle and naturally depends on both the return and volatility. With this methodology, the portfolio weight in the risky asset is inversely proportional to the recent squared volatility, which is a similar to the assumption in Moreira and Muir (2017). Intuitively, by avoiding high-volatility times the investor avoids risks, but if the risk-return trade-off is strong one also sacrifices expected returns, leaving the volatility timing strategy with no edge. Commonly, the volatility estimator is the realized volatility for the past few months. We propose a forward-looking model-based volatility estimate. The results show that the benefits from volatility-timing are enhanced when using this proposed measure for volatility.

Reward-risk timing is the combination of both return- and volatility-timing. Return-timing can be profitable with superior forecasting ability, yet ignoring the risk associated with a high return, for instance, would lead to poor risk-adjusted performance. The incorrect forecasts are not mitigated by their risk. On the other hand, volatility-timing is advantageous if the risk is not compensated fully by the reward, yet there may be cases when in fact the reward overcompensates the risk. Timing the market with both the expected return and volatility addresses the drawbacks of these individual approaches. The role of machine learning is to provide more accurate estimates by taking advantage of complex non-linear relationships between market variables and help make optimal decisions. With this, we provide a novel unifying framework for return- and volatility-timing as well as machine learning in portfolio allocation. 

An outline of the paper follows. Section \ref{sec:lit} reviews the literature. Section \ref{sec:headings} describes the portfolio allocation methodology, including the utility-maximization problem and models. Section \ref{sec:Empirical Results} demonstrates the results of using the machine learning portfolio allocation strategy, and Section \ref{sec:conclusion} concludes.

\section{Literature} \label{sec:lit}
Abundant work can be found on two strands of market timing, via expected returns and volatilites. Work can also be found on approaches combining the two, yet none to our knowledge integrate machine learning.

There is a long literature on expected-return timing. Kandel and Stambaugh (1996) examine equity return predictability and find that the optimal stock-versus-cash allocation can depend importantly on a predictor variable such as the dividend yield. Goyal and Welch (2008) comprehensively examine the performance of variables that have been suggested by the academic literature to be good predictors of the equity premium and find contradictory results. Johannes et al.\ (2014), however, find strong evidence that investors can use predictability to improve out-of-sample portfolio performance provided they incorporate time-varying volatility and estimation risk into their optimal portfolio problems.

There has also been a sizable interest in volatility-timing. Moreira and Muir (2017) showed volatility-managed factors outperform their buy-and-hold counterparts, modeling the optimal weight as a constant over the realized volatility for the previous month. Fleming et al.\ (2001) discussed the economic value of volatility timing, and Moreira and Muir (2019) found that investors who volatility time earn 2.4\% more annually than those who do not. Numerous papers have been written in response. Liu et al.\ (2019) found that the strategy in Moreira and Muir (2017) is subject to look-ahead bias since they choose the constant based on the full sample and that it is not easy to outperform the market with volatility timing alone. One finding in this paper is that simply replacing the constant with the expanding estimate of the unconditional mean excess return, which stays close to the constant chosen by Moreira, leads to similar performance\footnote{Our weight is constrained by a 150\% leverage limit so the alphas are not the same in the main results.}. % Another criticism of Moreira and Muir's paper raised by Cederburg et al.\ (2019) is that a strategy that has a positive alpha will not necessarily add to the investment value for an investor; the investment value increases only if the strategy yields a greater Sharpe ratio or higher investor utility when it is combined with the market or the investor’s existing portfolio.

Our main aim is to simultaneously perform expected return- and volatility-timing. Marquering and Verbeek (2004) study the economic value of predicting stock index
returns and volatility. They find that using simple linear models can lead to economically profitable performance in the monthly sample from 1970 to 2001. Our period is more recent. Also, Johannes et al.\ (2014) find statistically and economically significant out-of-sample portfolio benefits for an investor who uses models of return predictability when forming optimal portfolios, if accounting for estimation risk and allowing for time-varying volatility. We study a similar problem as these authors, however, not only with typical regression-based approaches but with machine learning models. 

Kirby and Ostdiek (2012) develop volatility- and reward-risk-timing strategies for the portfolio with many assets. Our paper considers the problem for the risk-free asset and the market while applying machine learning.

Gu et al.\ (2020) showed the benefit from using machine learning for empirical asset pricing, tracing the predictive gains to the allowance of non-linear predictor interactions. Trees and neural nets were the most successful in predicting returns. 

An article by Nystrup et al.\ (2016) proposes dynamic asset allocation using Hidden Markov Models that is based on detection of change points without fitting a model with a fixed number of regimes to the data, without estimating any parameters, and without assuming a specific distribution of the data. Our machine learning approach also does not assume a number of regimes, yet it does not discretize the portfolio weights.

To our knowledge, this is the first paper written on a machine learning approach to simultaneous return- and volatility-timing.

\section{Methodology}
\label{sec:headings}
We perform two tasks with machine learning that give the weight of the market index in our portfolio. First, we predict the market excess return next month with well-known macroeconomic and financial variables. Second, we estimate the prevailing volatility with a similar set of predictors. The weight of the equity index is proportional to the expected excess return and inversely proportional to the squared volatility estimate. The initial data the excess return and volatility models are trained on are from 1927 to 1957. The strategies are then optimized on out-of-sample data from 1958 to 1988 in a procedure called validation. Each month, the training data grows by one past observation and the models are refit. One set of models for each hyperparameter combination is kept. We select the combination of hyperparameters for Random Forest and Elastic Net that attains the highest predictive accuracy measured by $R^2$ over this validation period. Then the Random Forest and Elastic Net strategies are tested on a holdout set from 1989 to 2019, data that provides a final estimate of the models' performance after they have been validated, to prevent against backtest-overfitting (Bailey et al., 2015)\footnote{Holdout sets are never used to make decisions about which algorithms to use or for improving or tuning algorithms. Therefore, the performance on the holdout set is indicative of investment performance if an investor starts trading with the models and strategy today.}. Only one attempt on the holdout set is made. The general portfolio allocation approach is the following. For each month, update the machine learning models with the data only before that month, forecast the excess return and the volatility, and recompute the optimal weights. This gives us a time series of out-of-sample forecasts, portfolio returns, and corresponding performance metrics.

Using these time series, comprehensive summary statistics are computed to summarize the model and portfolio performance. We also conduct an array of tests to evaluate the robustness of our results. A key result is that the typical investor can benefit from reward-risk timing even if subject to realistic transaction costs and tight leverage constraints. A comparison of the Sharpe ratios and certainty equivalent (CE) yields of similar strategies that do not employ machine learning finds less impressive performance. Furthermore, examining the results of a series of time-series regressions gives evidence for positive alphas even after applying realistic transaction costs. %::Finally, we derive the theoretical alpha generation process to help explain these findings.:: 
The next section establishes the optimal trading rules followed.

\subsection{Portfolio Allocation} \label{portallocation}

%Most models of portfolio allocation with exact, closed-form solutions assume expected returns or stochastic volatility evolve continuously through time, a constant investment opportunity set, or single-period optimization. Our problem is harder due to the presence of time-varying risk premia and volatility across a discretized time horizon with periodic rebalancing. To find tractable solutions that are applicable to real-life investors, one can first consider the static one-period problem in Merton (1969) and Samuelson (1969), followed by stylized cases with time-varying expected returns and volatility which give our optimal weights.

Consider a power utility investor of terminal wealth $W_{t+\Delta t}$. 
\begin{equation}
{ \large U(W_{t+\Delta t}) = \frac{W_{t+\Delta t}^{(1-\gamma)} - 1}{1-\gamma}  } \normalsize,
\end{equation}
\noindent where $\gamma > 0$ is the coefficient of relative risk-aversion and as $\gamma \xrightarrow{} 1$, $U(W_{t+\Delta t}) = \ln W_{t+\Delta t}$. The investment universe with a risky asset with time-varying mean and variance and riskless asset constrained by a budget is defined by
\begin{gather}
%\begin{align}
r_t = \mu_t + \sigma_t\cdot z_t \\
%\end{equation}
%\begin{equation}
W_{t} = W_{t-1} \left (w_t\cdot \exp(r_t) + (1-w_t)\cdot \exp(r^f_{t}) \right ),
%\end{align}
\end{gather}

\noindent where $\mu_t$ is the expected log return on the risky asset, $\sigma_t$ is the time-varying volatility, $z_t$ is a standard normal random variable, $W_t$ is the investor's wealth at time $t$, $r_t^f$ is the risk-free asset log return, and $w_t$ is the portfolio weight in the risky asset at time $t$. In general form, we describe an asset’s excess return and volatility as additive prediction error models: 
\begin{gather}
%r_t = \mu_t + \sigma_t \cdot z_t \\
\mu_t = g_t \left ( \vec{x}_{t-1} \right ) + \epsilon_t \\
\log(\sigma_t^2) = h_t \left ( \vec{v}_{t-1}  \right ) + s_t,
\end{gather}
\noindent where $\vec{x}_{t-1}$ is the vector of predictor variables for the excess return model, $\vec{v}_{t-1}$ is the vector for the volatility model, $\epsilon_t$, and $s_t$ are potentially correlated normal random variables, $E[\epsilon_t|\mathcal{F}_{t-1}] = 0$, and $E[s_t|\mathcal{F}_{t-1}] = 0$. Functions $g_t$ and $h_t$ are to be estimated and can be non-linear. The well-known optimal weight\footnote{The derivation is shown in the appendix.} is 
\begin{equation} \label{eq:4}
    w_t^* = \frac{E[R_t-R_t^f|\mathcal{F}_{t-1}]}{\bar{\gamma} \cdot var[R_t |\mathcal{F}_{t-1}] },
\end{equation}
where $R_t=\exp(r_t)-1$.

With this portfolio allocation framework in mind, we examine a number of different variations of reward-risk timing for the utility-maximizing investor. One strategy is reward-risk timing with an expanding window estimate of the expected return and the last month's realized volatility as the prevailing volatility, referred to as the 'base' strategy. The investor relies on volatility clustering and has a simple estimate for the excess market return at time $t$. Specifically, in this strategy, volatility is computed from the daily returns for the past month but the risk premia with the full monthly sample until time $t-1$. The strategy's weights on the index are given by $\frac{1}{t-1}\sum_{i=1}^{t-1} (R_t-R_t^f) / (\bar{\gamma} \cdot \sigma_{t-1}^2 )$, a simple estimate of the optimal weight. Our conditionally mean-variance efficient or optimal reward-risk timing strategies employ machine learning and standard linear models models to 1) forecast the expected excess return for the next month with macroeconomic and financial variables and 2) estimate next month's volatility with a similar set of variables. Lastly, trading rules are examined that only use the return or volatility model forecast, with the other factor from the base strategy.

Our results support that machine learning models give more accurate estimates of the expected return than the simple unconditional mean, and the volatility estimates relative to the last month's realized volatility are similarly enhanced. We employ eleven macroeconomic and financial predictors for all the statistical models following the variable definitions detailed in Goyal and Welch (2008), including the dividend-price ratio (dp), earnings-price ratio (ep), book-to-market ratio (bm), net equity expansion (ntis), Treasury-bill rate (tbl), term spread (tms), default spread (dfy), inflation (infl), the high-quality corporate bond rate (corpr), long term rate of return (ltr), and stock variance (svar). An additional variable is the one-month lagged excess return. Lastly, for the expected return models, we also use one- to three-month lags of an enhanced measure of the payout yield from Boudoukh et al.\ (2007). Likewise, for the volatility models one- to three-month lags of the realized squared monthly volatilities are included.

Given accurate estimates of the two moments, the reward-risk timing strategies are able to avoid investing during most periods of low market reward and high risk. It is not surprising that even the performance of the simple reward-risk timing strategy is better relative to the buy-and-hold given that it is an extension of the risk-managed portfolio literature discussed in the next subsection. It has been shown that only using the recent volatility as a proxy for the near-future forecast has utility benefits (Moreira and Muir, 2019). The strategies employing machine learning, however, achieve the best results. Next, we look more closely at the volatility-timing strategy in the literature and the modification that is made to arrive at the base reward-risk timing strategy.

\subsubsection{Volatility-Timing}

\begin{figure}%[h] 
\captionsetup{labelfont=bf,font=small}
  \centering
  \includegraphics[scale=.7]{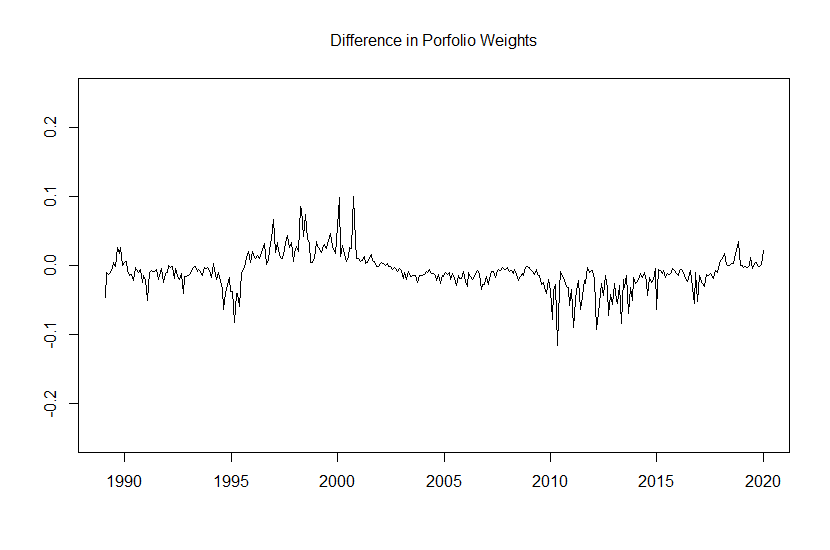}
  \caption{\textbf{Volatility-timing with a constant versus the expanding window estimate of excess return.} The constant $c$, which gives the volatility-timing strategy the same ending standard deviation as the buy-and-hold, over last month's realized volatility is plotted less the weight with an expanding excess return mean and a risk-aversion coefficient $\bar{\gamma}=6$. }
  \label{fig:cs}
\end{figure}

Moreira and Muir (2017) examine a volatility-managed portfolio constructed by scaling the portfolio weight of the market or factor $w_t$ by the inverse of the past month’s realized daily return variance. The strategy is motivated by their observation that changes in volatility over time are not offset by proportional changes in returns. The authors find that this volatility-timing strategy improves investment performance relative to the original market index and a wide range of asset pricing factors by reducing risk exposure when volatility is high (Liu et al., 2019). In this volatility-managed portfolio, the weight in the index is inversely proportional to the squared realized volatility,
\begin{equation}
{\large  w_t =\frac{c}{\hat{\sigma}_{t-1}^2}},
 \normalsize
\end{equation}
\noindent where $c$ is a constant and $\hat{\sigma}_{t-1}^2$ is the realized return variance in month $t - 1$. $\hat{\sigma}_{t-1}^2$ is computed from the 22 average daily returns over the month
\begin{equation}
\hat{\sigma}_t^2(f) = RV_t^2(f) = \sum_{d=1/22}^1 \left (f_{t+d}^D - \frac{\sum_{d=1/22}^1 f_{t+d}^D}{22} \right )^2, 
\end{equation}
where $f^D$ is the daily excess return. The constant $c$ is set in Moreira and Muir (2017) such that the strategy's standard deviation matches that of the buy-and hold for ease of interpretation. Liu et al.\ (2019) point out that choosing $c$ based on the unconditional volatility over the entire period is an in-sample approach and is thus subject to look-ahead bias. While this is correct, simply using the historical average excess return instead of the constant gives similar weights and performance over time. This is not surprising since the historical mean divided by the commonly used risk-aversion coefficient $\bar{\gamma} = 6$, for instance, produces a numerator that stays consistently close to the exact value of $c$, the constant which makes the standard deviation of the volatility-managed strategy equal to that of the buy-and-hold\footnote{Because our data has a different sample period, the value here does not match that in the papers above.}. Figure \ref{fig:cs} shows the effects on the portfolio weights and plots the difference between the two weights $c/\hat{\sigma}_{t-1}^2$ and $\frac{1}{t-1}\sum_{i=1}^{t-1} (R_t-R_t^f) /(6 \cdot \hat{\sigma}_{t-1}^2)$ from 1989 to 2019.  

%\end{multicols}

%\begin{multicols}{2}

\noindent The two weights stay close to each other over the period and the difference generally does not exceed 10\% in absolute value. 

The discussion above provides an intuition for why this modified version of volatility-timing, or base reward-risk timing, achieves investment performance for the market portfolio similar to volatility-timing in Moreira and Muir (2017). The results are discussed in Section \ref{sec:Empirical Results}. To come to the full strategy, we first look at the standard linear and machine learning models in the next sections.

\subsection{Elastic Net}  \label{ARMA}

Starting with a standard linear model,
\begin{equation}
     y_t = \mu + \sum_{i=1}^{m} \beta_i x_{i,t-1} + \epsilon_t
\end{equation}
where $y_t$ can be either the log excess return $r_t-r_t^f$ or the volatility $\sigma_t$, we can consider various forms of regression regularization to deal with the high dimensionality of the predictor set. This gives alternate procedures to estimate the model coefficients from OLS. First we describe LASSO, penalized regression that is designed to prevent overfitting with shrinkage.
 
To fit a model, minimize the objective function
\begin{equation}
    \min_{\mu,\beta_1,...,\beta_m } \frac{1}{T} \sum_{t=1}^{T} \left (y_t - \mu - \sum_{j=1}^{m}\beta_j x_{j,t-1} \right )^2 + \lambda \sum_{j=1}^{m} |\beta_j|,
\end{equation}
where $\lambda \geq 0$ is the shrinkage parameter on the $l_1$ penalty. A higher value of $\lambda$ places a higher penalty on the coefficients' absolute values, selectively shrinking them, and a high enough $\lambda$ can make coefficients zero. This produces a looser fit on the training data but less chance of over-fitting in terms of out-of-sample forecasts. Setting $\lambda = 0$ gives the same coefficients as OLS. To select the optimal value, validation is typically done by testing the performance for a range of values on an out-of-sample data set. The parameter value that gives the maximum predictive accuracy is then used in the model on a distinct out-of-sample set for which results are reported.

While the LASSO fitting method typically improves predictions relative to the OLS model, it can sometimes select one predictor arbitrarily from a group of correlated predictors. Zou and Hastie (2005) proposed Elastic Net, regression with both $l_1$ and $l_2$ loss, which adds a second parameter and makes variable selection more robust. The objective function is
\begin{equation}
    \arg_{\mu,\beta_1,...,\beta_m } \frac{1}{T} \sum_{i=1}^{T} (y_t - \mu - \sum_{j=1}^{m}\beta_j x_{j,t-1} )^2 + \lambda ( \alpha \sum_{j}^{m} |\beta_j| + \frac{1}{2} (1-\alpha) \sum_{j=1}^{m} \beta_j^2 ) .
\end{equation}
The parameter $0 \leq \alpha \leq 1$ controls the blending of the $l_1$ and $l_2$ loss. Using $\alpha > 0$ results in a stronger tendency to select groups of correlated predictors. The parameters $\alpha$ and $\lambda$ for the Elastic Net model are chosen with the sample from 1958 to 1988 with cross validation as described in Section \ref{sec:headings}. The out-of-sample predictions for LASSO or Elastic Net are given by
\begin{equation}
     \hat{y}_{t+1} = \hat{\mu} + \sum_{i=1}^{m} \hat{\beta_i} x_{i,t} .
\end{equation}
The predictions, like for a standard OLS linear model, are a weighted sum of variables. The next subsection discusses the machine learning model Random Forest, which relies on recursive partitioning of the feature space to make predictions, and why it can perform better than linear models in our portfolio allocation problem. 

\subsection{Random Forest}  \label{Random Forest}
Random Forest is an ensemble machine learning algorithm developed by Breiman (2001). The prediction by a Random Forest model is the majority vote across all the individual decision tree learners (Hastie et al., 2017). The default tree bagging procedure draws $B$ different bootstrap samples of the training data and fits a separate classification tree to the $bth$ sample. The forecast is the average of the trees' individual forecasts. Trees for a bootstrap sample are usually deep and overfit, meaning each has low bias but is inefficiently variable. Averaging over the $B$ predictions reduces the variance and stabilizes the trees’ forecast performance. Algorithm 2 gives the procedure used to construct a Random Forest with the implementation by Liaw and Wiener (2002).

\medskip
%\end{multicols}
\begin{algorithm}%[H] \label{alg:2}
%\begin{varwidth}[t]{0.4\textwidth}
%\SetAlgoLined
\KwResult{The ensemble of trees \{$T_b\}^B$ }

 \For{$b=1$ to $B$}{
  \begin{enumerate}
  \item Draw a bootstrap sample $\mathbf{Z^*}$ of size $n$ from the training data.
  
  \item Grow a random-forest tree $T_b$ to the bootstrapped data by \\ recursively repeating the following steps for each terminal node \\ of the tree, until the minimum node size fraction $s_{min}$ or the maximum \\ number of terminal nodes $k_{max}$ are reached.
  \begin{enumerate}
    \item Select $m$ variables at random from the $p$ variables
    
    \item Pick the best variable/split-point among the $m$.
    
    \item Split the node into two child nodes.
    \end{enumerate}
 \end{enumerate}
  
}
 \caption{Random Forest}
% \end{varwidth}
\end{algorithm}
%\begin{multicols}{2}
\medskip
\noindent The prediction at a new point, $\vec{x}$, is 
\begin{equation}
    \hat{f}(\vec{x}) = \frac{1}{B} \sum_{b=1}^B \hat{T}_b(\vec{x}) ,
\end{equation} the average of all the individual trees' predictions. 

Random forests give an improvement over bagging with a variation designed to reduce the correlation among trees grown from different bootstrap samples. If most of the bootstrap samples are similar, the trees trained on these sample sets will be highly correlated. The average estimators of similar decision trees do not perform much better than a single decision tree. If, for example, among the variables, last month's dividend yield is the dominant predictor of the return, then most of the bagged trees will have low-depth splits on the most recent yield, resulting in a large correlation among their predictions. Trees are de-correlated with a method known as "random subspace" or "attribute bagging," which considers only a random subset of $m$ predictors out of $p$ for splitting at each potential branch. In the example, attribute bagging will ensure early branches for some trees will split on predictors other than the most recent dividend yield. Since each tree is grown with different sets of predictors, the average correlation among trees further decreases and the variance reduction relative to standard bagging is larger (Gu et al.\, 2020)\footnote{Because this makes Random Forest a non-deterministic algorithm, we average the results for multiple different seeds.}. The number of variables randomly sampled as candidates at each split, $m$, the number of bootstrap samples, $B$, the minimum fraction of observations in the terminal nodes, $s_{min}$, and $k_{max}$ are the tuning parameters optimized with validation. A detailed algorithm for classification trees can be found in the Appendix.

The parameters $m$, $s_{min}$, $k_{max}$, and $B$ are tuned with the sample from 1958 to 1988. To test against parameter over-fitting, the final values are kept on the holdout time period from 1989 to 2019, for which results are reported, and only one attempt is made on the period. 

\subsubsection{Why Apply Random Forest to Portfolio Allocation?}

With an understanding of the Random Forest model, we can discuss why this it is preferred over alternative machine learning methods for this portfolio allocation problem.

Tree-based learning models like Random Forest have certain desirable characteristics such as being non-metric, meaning there are no inherent assumptions of distributions in data. Decision trees are also scale invariant; rescaling the features by nonzero numbers do not change their predictions. The number of parameters typically optimized in Random Forest is fewer than many other machine learning models. Deep neural networks, for example, can have hundreds of parameters to estimate, and the possible configurations of hidden layers and neurons are practically uncountable. 

There is also the problem that financial data are notoriously noisy. Risk premia are difficult to forecast as market efficiency diminishes the signal-to-noise ratio in well-known variables. The risk premia estimation problem is further complicated by potential shifts in the data distributions. If a model mostly relies on idiosyncratic relationships in past data, the out-of-sample performance will significantly suffer when those patterns fade over time. Random Forest can both find complex signals and mitigate the effect of changing relationships between predictors and the target variable such as excess returns with the random subspace method. If one tree is grown to capture the relationship between expected returns and the inflation and term spread variables, the tree may accurately predict the expected excess returns in some market environments, but not in all. In certain time periods, the dividend yield, for instance, may be more strongly correlated with excess market returns. Since Random Forest grows many trees with different variables, if there are changes in the data distributions, some of the trees might not perform well, but the results of the forest should largely remain unchanged. In other words, while a single tree may capture the relationships in the training data well, it is less stable. In general, a forest model can be used to reduce the effect of noisy data.

%We show how Random Forest mitigates this effect with a small example. %Consider August, 2008, the month before the two worst monthly returns for the market during the Great Financial Crisis. The first tree in our Random Forest model, trained on all observations from is shown 

\subsection{Conditional Excess Return and Volatility Estimation}

Forecasting individual stock returns is explored extensively in Gu et al.\ (2020). We focus on the aggregate market excess returns, yet the general methodology for both excess returns and volatility could be used on specific stocks too. This is left as a subject for future research. 

For optimal portfolio construction, the weight of the market index should increase when the investor expects a greater excess return, holding all else constant. To estimate the excess return each month, we borrow from the standard literature which commonly employs the variables from Goyal and Welch (2008). Additionally, we use a variation of lagged dividend yields as predictors. The importance of the dividend yield in the allocation is robust to the "data-mining" consideration (Kandel and Stambaugh, 1996), and it has been shown to explain equity return predictability in Johannes et al.\ (2004) for example. In traditional theory, the dividend yield can explain equity prices since prices are the discounted future cash flows. Boudoukh et al.\ (2007) research a measure of net payout yield incorporating both share repurchases and issuances which, compared to dividend yields, can have a stronger association with returns as firms have shifted the ways they distribute earnings to their shareholders. We use net payout yields in lieu of traditional dividend yields and, in line with previous findings by Boudoukh et al., observe better predictive ability in the linear and machine learning models. Higher order lags of the payout yield up to three months still contain valuable information. In traditional literature, a higher past month's dividend yield is indicative of a higher chance of a positive excess return (Fama and French, 1988). Yet the yield two months ago still has information about the overall trend in the market. We trace the predictive gains of our approach to the presence of interaction effects between payout yields at different months and the other macroeconomic and financial variables, which Random Forest can detect.

%Lagged risk-free rates are also used as predictors. Relatively slow-moving, the rates serve mostly as indicators of the current market regime. Their inclusion magnifies the predictive performance of the linear and machine learning models. 
A feature of our approach is the exclusion of outliers. We omit the top decile of returns in absolute value, with this cutoff best performing in the validation set. We use the same cutoff in the test period. While trimming achieves better predictive accuracy, one could point out that this may limit the return model's ability to identify extreme market events. Our second model, however, which forecasts volatility is better at anticipating months with extreme market conditions. It can be seen that in combination the two models balance each other and improve portfolio performance.

Volatility has a central role in optimal portfolio selection, derivatives pricing, and risk management. These applications motivate an extensive literature on volatility modeling. Starting with Engle (1982), researchers have fit a variety of autoregressive conditional heteroskedasticity (ARCH), generalized ARCH (Bollerslev, 1986), and stochastic volatility models to asset returns (Fleming et al., 2001). GARCH models are widely used for their ability to permit a wide range of behavior, in particular, more persistent periods of high or low volatility than seen in an ARCH process (Ruppert and Matteson, 2015). We model the volatility as a function of macroeconomic and financial variables as well as past realized volatilities.   

We use the variables described in Section \ref{portallocation}. The realized daily return variance for a month is given by Eq. 8. The variance is highly persistent, as using simply the previous month's is sufficient for an out-of-sample nearing 50\% $R^2$. Employing lagged realized volatilities as predictors in our machine learning models achieves even higher accuracies.

\section{Empirical Results} \label{sec:Empirical Results}

\subsection{Data Description}

This paper uses monthly time series from Kenneth French’s\footnote{\url{http://mba.tuck.dartmouth.edu/pages/faculty/ken.french/data_library.html}} website on the market return (Mkt) and risk-free asset return (Rf) from 1927 to 2019, with 1927-1957 as the initial training period, 1958-1988 the validation period, and 1989-2019 the test period. Daily returns are retrieved to compute the realized volatilities. The monthly data for the conditioning factors are from Amit Goyal's website\footnote{\url{http://www.hec.unil.ch/agoyal}}.

The payout yield data is from Michael Robert's website\footnote{\url{http://finance.wharton.upenn.edu/~mrrobert/styled-9/styled-13/index.html}}, which is updated to cover January 2011 to December 2019 and is derived from all firms continuously listed on the NYSE, AMEX, or NASDAQ indices. For the updated data, CRSP monthly data at the firm-level and the same aggregation procedure to form the payout yields as by Boudoukh et al.\ (2007) is used. This payout yield is a more inclusive measure of total payouts than standard dividend yields and is achieved via the ‘net payout’ of Boudoukh et al.\ (2007). It includes share issuances and repurchases in addition to the traditional cash dividend yields. In recent years share repurchases have played a more important role in total payouts to shareholders. For example, Boudoukh, Richardson, and Whitelaw (2006) report a significantly higher forecast $R^2$ when using various measures of the payout yield (i.e. including repurchases) than the dividend yield. 

\subsection{Predictive Performance}

To assess the predictive performance for the simple, linear, and machine learning models, we measure their out-of-sample $R^2$ and directional accuracies. The out-of-sample $R^2$ for excess returns is calculated as
\begin{equation}
    R^2_{os} = 1 - \frac{\sum_{t\in \mathcal{T} } (f_{t+1} -\widehat{f_{t+1}} )^2  }{\sum_{t\in \mathcal{T} } (f_{t+1} -\overline{f_{t+1}} )^2 }
\end{equation}
where $\mathcal{T}$ denotes the set of points not used for model training and $f$ are the monthly market excess returns, $\hat{f}$ are the model forecasts, and the mean excess return $\overline{f_{t+1}}$ is the competing forecast. The notation for excess return, $f$, is for readability and also reflects that reward-risk timing can be applied to factors other than the market. The $R^2$ for the volatility models is computed in the same way. 

Table \ref{tab:pred} contains the $R^2$ values and directional accuracies for each forecasting model for excess returns and volatility.

\begin{table}[!htbp] 
\captionsetup{labelfont=bf,font=normalsize}
\caption{\textbf{Out-of-Sample Forecasting Accuracy}}
\justifying{\small{\noindent In this table are the out-of-sample $R^2$ and directional accuracies from 1989 to 2019 for the various excess return and volatility models. The directions for volatility are based off the mean. }}

\medskip

\centering
\label{tab:pred}
\begin{tabular}{lccc} \hline
        Model & $R^2$ & Directional Accuracy (\%)  \\ \hline
         & Excess Returns &   \\ \hline
Prevailing Mean     & -0.0012 & 64.25 \\
Linear Model        & -0.0351 & 63.17 \\
Elastic Net         & -0.0273 & 63.17 \\
Random Forest       & 0.0052  & 64.52 \\ \hline
 & Volatility &   \\ \hline
Previous Realized Volatility & 0.4437  & 78.49 \\
Linear Model        & 0.5469  & 79.84 \\
Elastic Net         & 0.5451  & 80.65 \\
Random Forest       & 0.5008  & 80.91 \\ \hline
\end{tabular}
\end{table}

\noindent Random Forest is the best performing method for excess returns and attains the only positive $R^2$, 0.52\%, and correctly identifies the correct sign of the excess return 64.52\% of the time. The expanding window mean estimate is slightly negative, and the linear models including Elastic Net do not beat the simple mean. For the Random Forest excess return models the optimal values we find for $s_{min}$, $k_{max}$, the number of trees, and the number of variables to select from at each split ($m$) are 0.95, 2, 500, and 4, respectively. The parameters have varying degrees of influence on the model. The larger the value of $s_{min}$ the more shallow the trees will be in general, lessening the chance to overfit. The excess return data sets have significant noise so a large value is not surprising. Generally, once a sufficient number of trees has been reached tuning is not necessary. The maximum number of terminal nodes also controls the depth of trees, but more directly. Reducing $m$ reduces the correlation between trees. The shrinking parameter $\lambda$ is 0.07 and blending parameter $\alpha$ is 0.1 for Elastic Net.

For volatility forecasting, the linear model and elastic net attain the highest $R^2$ values of 54.69\% and 54.51\%, respectively. Random Forest produces an $R^2$ above 50\% as well and the highest directional accuracy of 80.91\%. For the volatility Random Forest models, the respective values for $s_{min}$, $k_{max}$, the number of trees, and the number of variables to select from at each split ($m$) are 0.01, 12, 500, and 4. The trees are grown much deeper than for the excess return models as volatility is more predictable. For Elastic Net, $\lambda$ is 0.3 and $\alpha$ is 0.1.

%The non-parametric test proposed by Henriksson and Merton (1981) is asymptotically equivalent to a one-tailed test on the significance of the coefficient $\beta$ in the regression
%\begin{equation}
%   \hat{f}_{t+1} = \alpha + \beta f_{t+1} + \epsilon_{t+1}.
%\end{equation}

Next, we show variable importance for the models measured by estimated Shapley values (Shapley, 1953). We use an algorithm called Kernel SHAP to approximate the values (Lundberg and Lee, 2017). SHAP calculates the impact of each feature on the predictions made by the learned model. Given an input vector $\vec{x}$ and a trained model $f$, SHAP approximates $f$ with a simple model $g$ that can easily explain the contribution of each feature value. The Kernel SHAP algorithm involves the following steps:
\begin{enumerate}
    \item Sample $S$ coalitions $\vec{z_k} \in \{0,1\}^M$ from $2^M-2$ total, where $M$ is the number of variables, 0 indicates a variable is absent, and 1 indicates it is present\footnote{We use a sample of 1,000 cases.}.
    \item Convert $\vec{z_k}$ into the original space by replacing absent feature values with either sampled or reference values\footnote{We use variable means as the reference values.} and compute predictions $f(h(\vec{z_k}))$ for each sampled $\vec{z}$, where $h$ is the conversion function.
    \item Fit the weighted linear model $g(\vec{z}) = \phi_0 + \sum_{j=1}^M\phi_j z_j$ by minimizing $L = \sum_{\vec{z} \in Z} [ f(h(\vec{z}) - g(\vec{z}) ]^2 \pi(\vec{z})  $, where $\pi(\vec{z}) = (M-1) /( {M\choose {\vert\vec{z}\vert} } \vert \vec{z} \vert (M - \vert \vec{z} \vert ))$ is the SHAP kernel.
    \item Return estimated Shapley values $\phi_j$, the coefficients from the linear model.
\end{enumerate}

\begin{figure}%[h]
\captionsetup{labelfont=bf,font=small}
  \centering
  \includegraphics[scale=.7]{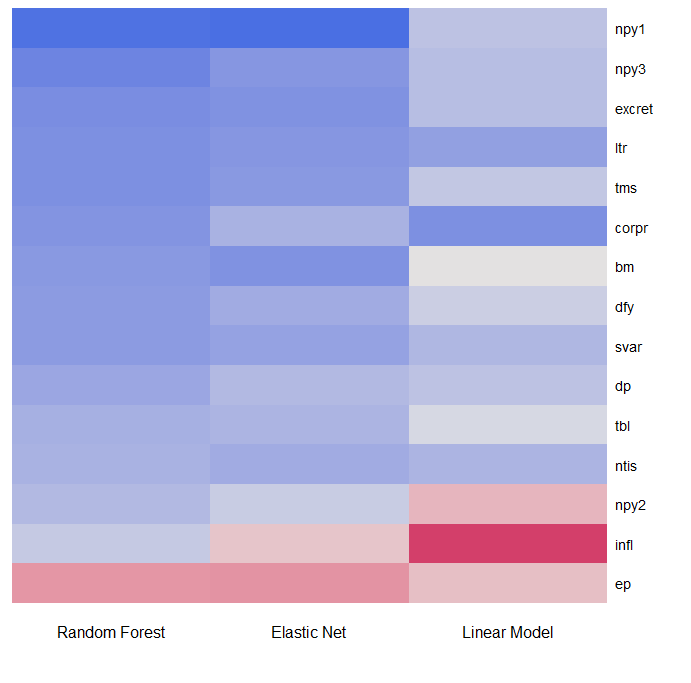}
  \caption{\textbf{SHAP values for excess return factors.} This figure shows the overall importance of factors for the excess return models for the 1989 to 2019 period sorted by most positive positive contribution for Random Forest. Variable names are defined in Section 3.1. npy indicates net payout yield and the number refer to the order of the lag.}
  \label{fig:figshapret}
\end{figure}

\begin{figure}%[h]
\captionsetup{labelfont=bf,font=small}
  \centering
  \includegraphics[scale=.7]{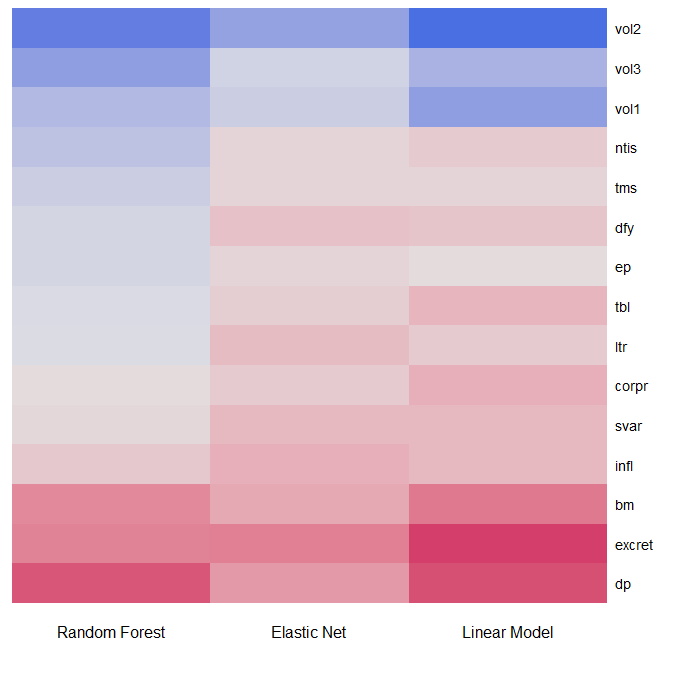}
  \caption{\textbf{SHAP values for volatility factors.} This figure shows the overall importance of factors for the volatility models for the 1989 to 2019 period sorted by most positive contribution for Random Forest. Variable names are defined in Section 3.1. vol indicates monthly realized volatility and the number refer to the order of the lag.}
  \label{fig:figshapvol}
\end{figure}

In Figure \ref{fig:figshapret} we show the average factor contributions to the excess return predictions over the 1989 to 2019 period and in Figure \ref{fig:figshapvol} is the same for volatility forecasts. The darkest blue cells indicate the strongest positive factor contributions to an excess-return forecast, and the lightest cells mean the strongest negative factor contribution.

For the excess return Random Forest model, high values of net payout yield lagged by one and three months, excess return, and the long-term rate of return are the most indicative of a higher forecasted excess return, on average. Earnings to price is the top contributor for a smaller forecast. %Elastic Net and the OLS model find that the long-term rate of return and the two-month lagged payout yield are most important for a predicted positive return. 

The volatility Random Forest, Elastic Net, and Linear Models have the most recent realized volatilities and net issuance as the largest contributors for a higher forecasted volatility next month. The dividend-price ratio is the largest negative contributor for Random Forest.

With these forecasting characteristics in mind, we next discuss the risk-adjusted performance of the strategies and models.

\subsection{Risk-Adjusted Returns}

\begin{figure}%[h]
\captionsetup{labelfont=bf,font=small}
  \centering
  \includegraphics[scale=.7]{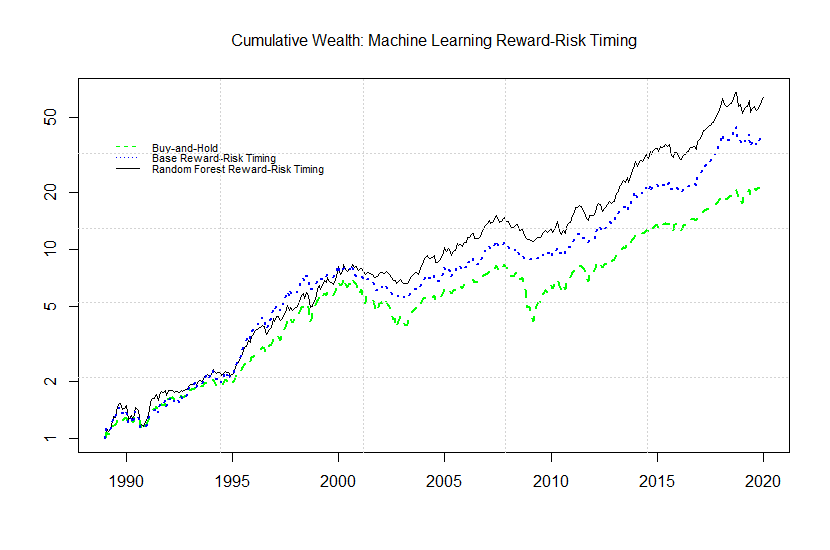}
  \caption{\textbf{Cumulative returns of reward-risk timing to market index (200\% leverage limit).} This figure plots the cumulative returns of the base reward-risk timing strategy in blue and Random Forest reward-risk timing in black against the market index in green from 1989 to 2019. The vertical axis is in log-scale.}
  \label{fig:fig2}
\end{figure}

\begin{figure}%[h]
\captionsetup{labelfont=bf,font=small}
  \centering
  \includegraphics[scale=.7]{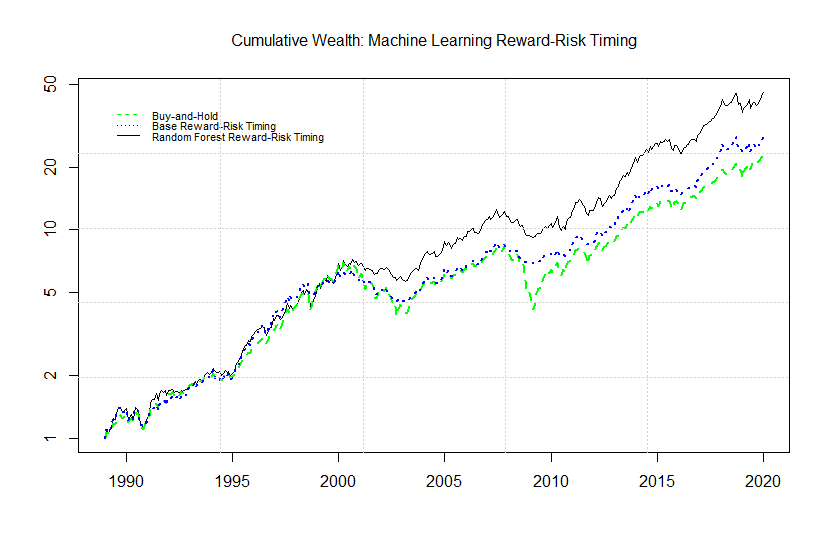}
  \caption{\textbf{Cumulative returns of reward-risk timing to market index (150\% leverage limit).} This figure plots the cumulative returns of the base reward-risk timing strategy in blue and Random Forest reward-risk timing in black against the market index in green from 1989 to 2019. The vertical axis is in log-scale.}
  \label{fig:fig3}
\end{figure}

This section discusses the out-of-sample investment performance for machine learning calibrated reward-risk timing and makes the relevant comparisons. We invest \$1 in the start of 1989 as an investor with a coefficient of relative-risk aversion $\bar{\gamma} = 4$ and plot the cumulative returns to each strategy on a log scale in Figures \ref{fig:fig2} and \ref{fig:fig3} without short-selling and with 100\% and 50\% leverage constraints, respectively\footnote{The figures and tables in this section are all with $\bar{\gamma} = 4$ except for Table 3. The results do not change significantly for other values.}. For the rest of the paper, we impose the more realistic portfolio constraint, preventing the investor from taking more than 50\% leverage as in Campbell and Thompson (2008): that is, confining the portfolio weight on the market index to lie between 0\% and 150\%.

%\end{multicols}
%\begin{multicols}{2}

The investments that reward-risk time realize relatively steady gains. The final wealth accumulates to around \$46 and \$28 at the end of the sample for the Random Forest and base (expanding sample mean reward estimate and previous month realized volatility risk estimate) strategies, respectively, versus about \$23 for the buy-and-hold. During stable periods of high market returns, the Random Forest models aptly forecast higher excess returns and lower volatility, leading to greater performance. The models also lead to better performance during recessions and avoid as high allocations as the passive strategy. The 'break-away' moment for Random Forest from the base reward-risk timing strategy is around 2000. During periods of market expansion, the Random Forest portfolio takes more risk which leads to steadily increasing outperformance relative to the base portfolio. Since all the model parameters are chosen with data before 1989, the results cannot be easily explained by the particular choice of machine learning model parameters.  

\begin{figure}%[h]
\captionsetup{labelfont=bf,font=small}
  \centering
  \includegraphics[scale=.7]{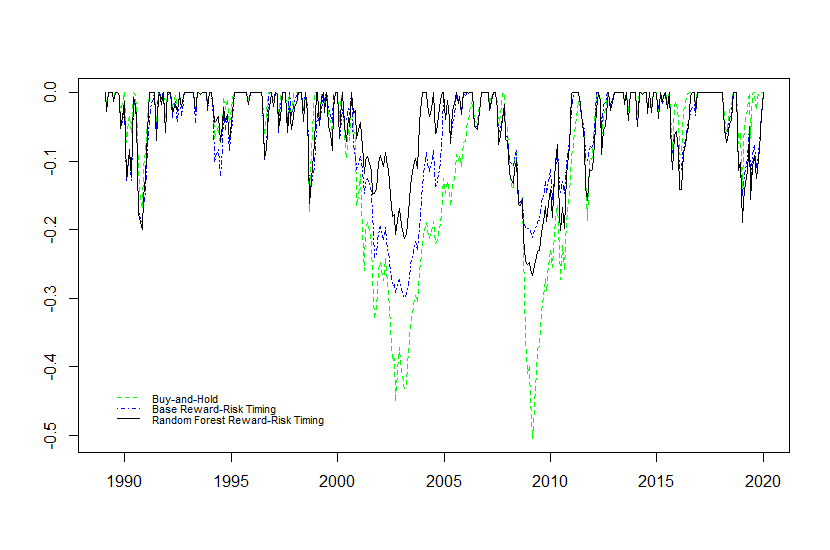}
  \caption{\textbf{Drawdowns of reward-risk timing to market index.} This figure plots the drawdown of the base reward-risk timing strategy in blue, machine learning reward-risk timing in black against the market index in green from 1989 to 2019.}
  \label{fig:fig5}
\end{figure}

It is also valuable to look at the drawdowns for the strategies. Figure \ref{fig:fig5} plots the drawdown starting from 1989 of the two strategies relative to the market, which helps us understand when our strategies lose money relative to the buy-and-hold. 
%\begin{multicols}{2}
%\end{multicols}
\noindent The base reward-risk strategy takes relatively less risk when volatility is high (e.g., the 2000s) and thus, not surprisingly, it diminishes the largest markets losses concentrated in those times. The machine learning analog has a pattern of losses similiar to simple reward-risk timing, yet it diminishes the severity of many losses and to a high degree for some of the most extreme negative returns. For the sharp market losses starting in 2007, the first major drawdown, the Random Forest models' response is delayed, due to the very sudden drop. Yet for the other major drawdown in 2001, our Random Forest models are able to recognize the incoming negative returns because the drops are more staggered, cutting the losses felt by investors greatly. This is seen clearly in the Dot-com recession from 2000 to 2002, where using machine learning allows investors to more than halve losses during this time. In the last recession of 2007--2008, due to the extremely sharp onset, our return machine learning model reduces risk exposure slightly too late, yet the information in the volatility estimate still correctly steers market exposure down. Reward-risk timing never has a drawdown greater than 30\% of the portfolio value and greatly mitigates losses during severe recessions.

The risk-adjusted returns from machine learning portfolio allocation are substantially higher than simple reward-risk timing and the buy-and-hold. Table \ref{tab:2} displays the Sharpe ratios for each portfolio allocation strategy for the sample from 1989 to 2019. We run the trading rules on it with the same parameters and seeds as the 1958--1988 sample after they are finalized.

%\end{multicols}
\begin{table}[!htbp] 
\captionsetup{labelfont=bf,font=normalsize}
\caption{\textbf{Sharpe Ratios}}
\justifying{\small{\noindent In this table are the out-of-sample annual returns, standard deviations, and Sharpe ratios for the test period from 1989 to 2019 for the trading rules. Mkt denotes the buy-and-hold.}}

\medskip

\centering
\label{tab:2}
\resizebox{\textwidth}{!}{
\begin{tabular}{lccccccl}
\hline
Strategy                    & Annual Return (\%) & Standard Deviation (\%) & Sharpe Ratio & \multicolumn{1}{c}{} \\
\hline
% \multicolumn{5}{c}{ $\bar{\gamma} = 4$} \\ \hline
Mkt                      & 11.21 & 14.57 & 0.57 \\
Base                     & 11.63 & 13.03 & 0.67 \\
\\
Linear Model Optimal     & 13.84 & 16.30  & 0.67 \\
Linear Model Returns     & 13.09 & 15.40  & 0.66 \\
Linear Model Volatility  & 12.08 & 13.07 & 0.71 \\
\\
Elastic Net Optimal      & 14.01 & 16.10  & 0.69 \\
Elastic Net Returns      & 13.15 & 15.12 & 0.68 \\
Elastic Net Volatility   & 12.08 & 13.18 & 0.70  \\
\\
Random Forest Optimal    & 13.42 & 14.39 & 0.73 \\
Random Forest Returns    & 12.64 & 13.57 & 0.72 \\
Random Forest Volatility & 12.08 & 13.54 & 0.68 \\
       \hline  

\end{tabular}
}
\end{table}
%\begin{multicols}{2}

All the active strategies outperform the buy-and-hold on a risk-adjusted basis for the out-of-sample period. Reward-risk timing with Random Forest, using Random Forest for both conditional excess return and volatility estimates, gives the highest Sharpe ratio of 0.73, which is a 28\% increase from the buy-and-hold. An investor who reward-risk times with machine learning gains about 2 percentage points on return per year relative to passively investing, while decreasing the risk.

To quantify the economic relevance of our results and facilitate comparison, we consider the perspective of the power-utility investor. Table \ref{tab:3} contains the average monthly realized utilities, certainty-equivalent (CE) yields computed as the inverse utility function of the average realized utility, and terminal wealths for different risk aversion coefficients. 

\begin{table}[!htbp]
\captionsetup{labelfont=bf,font=normalsize}
\caption{\textbf{Average Realized Utilities}}

\centering
\label{tab:3}
\justifying{\small{\noindent In this table, the average monthly realized utilities, annual CE yields, and terminal wealths for each strategy are shown under risk aversion coefficients 4 and 6 for the 1989 to 2019 out-of-sample period.}}

\medskip
\centering
\begin{tabular}{llll}
\multicolumn{4}{c}{Power Utility Investor}     
\\ \hline                
\multicolumn{4}{c}{$\bar{\gamma} = 4$}  
\\ \hline
Strategy                 & Utility & CE yield & Terminal Wealth \\ \hline
Mkt                      & 0.0056  & 0.0677   & 22.8993         \\
Base                     & 0.0067  & 0.0813   & 27.8025         \\
\\
Linear Model Optimal     & 0.0068  & 0.0827   & 47.2390         \\
Linear Model Returns     & 0.0067  & 0.0816   & 39.2116         \\
Linear Model Volatility  & 0.0070  & 0.0856   & 31.9458         \\
\\
Elastic Net Optimal      & 0.0071  & 0.0863   & 50.3243         \\
Elastic Net Returns      & 0.0069  & 0.0845   & 40.5950         \\
Elastic Net Volatility   & 0.0070  & 0.0850   & 31.7764         \\
\\
Random Forest Optimal    & 0.0075  & 0.0917   & 45.5343         \\
Random Forest Returns    & 0.0073  & 0.0887   & 37.1885         \\
Random Forest Volatility & 0.0068  & 0.083    & 31.2777         \\

\\ \hline
\multicolumn{4}{c}{$\bar{\gamma} = 6$} 
\\ \hline
Mkt                      & 0.0036  & 0.0437   & 22.8993         \\
Base                     & 0.0055  & 0.0670   & 19.6713         \\
\\
Linear Model Optimal     & 0.0057  & 0.0690   & 35.8601         \\
Linear Model Returns     & 0.0053  & 0.0651   & 29.7270         \\
Linear Model Volatility  & 0.0056  & 0.0682   & 18.3737         \\
\\
Elastic Net Optimal      & 0.0059  & 0.0715   & 36.6955         \\
Elastic Net Returns      & 0.0056  & 0.0684   & 30.6198         \\
Elastic Net Volatility   & 0.0056  & 0.0678   & 18.1856         \\
\\
Random Forest Optimal    & 0.0058  & 0.0714   & 24.6255         \\
Random Forest Returns    & 0.0059  & 0.0723   & 25.3119         \\
Random Forest Volatility & 0.0056  & 0.0678   & 18.6655            

\\ \hline

\end{tabular}
\end{table}

For a risk-aversion coefficient $\bar{\gamma} = 4$, the certainty-equivalent (CE) yield for the Random Forest combined models is the highest at 9.17\%. The optimal Elastic Net and Linear and base strategies also give CE yields markedly greater than the market with 8.63\%, 8.27\%, and 6.77\%, respectively. The average monthly utility is also 33.9\% greater for Random Forest reward-risk timing than the buy-and-hold. For a smaller risk aversion $\bar{\gamma} = 6$, the utility increase is 61.1\%. In perspective, Campbell and Thompson (2008) estimate that the utility gain of timing expected returns is 35\% of lifetime utility. Reward-risk timing can generate larger gains relative to solely focusing on the reward or risk factors. 

Next, we run a series of time-series regression of the strategies on each other and the market index,
\begin{equation}
   f_{t+1}^a = \alpha + \beta f_{t+1}^b + \epsilon_{t+1} ,
\end{equation}
\noindent where $f_{t+1}$ are the monthly excess returns. A positive intercept implies that the strategy $a$ increases Sharpe ratios relative to strategy $b$. When this test is applied to systematic factors (e.g., the market portfolio) that summarize pricing information for a wide cross-section of assets and strategies, a positive alpha implies that our portfolio-allocation strategy expands the mean-variance frontier.

\begin{table}[!htbp]
\captionsetup{labelfont=bf,font=normalsize}
\caption{\textbf{Strategy Alphas}}

\centering
\label{tab:4}
\justifying{\small{\noindent In this table, we run time-series regressions of each strategy on the market and on one another $f_{t+1}^a = \alpha + \beta f_{t+1}^b + \epsilon_{t+1}$. The data are monthly and the sample period is 1989 to 2019. Standard errors are in parentheses and are adjusted for heteroskedasticity (White, 1980). The alphas and errors are annualized in percent per year by multiplying monthly values by 12.}}

\medskip
\centering
\begin{tabular}{llccccc}
\multicolumn{6}{c}{Univariate Regressions}     
\\ \hline                                
%& &  $\bar{\gamma} = 4$ & &  \\ \hline
$f_a$                                      & $f_b$                                      & Beta ($\beta$) & Alpha ($\alpha$) & $R^2$ & $N_{obs}$ \\ \hline
Random Forest Optimal & Mkt                  & \begin{tabular}[c]{@{}c@{}}0.57\\ (0.03)\end{tabular} & \begin{tabular}[c]{@{}c@{}}3.37\\ (1.42)\end{tabular} & 0.76 & 372 \\
Random Forest Optimal & Base                 & \begin{tabular}[c]{@{}c@{}}1.04\\ (0.03)\end{tabular} & \begin{tabular}[c]{@{}c@{}}1.44\\ (0.94)\end{tabular} & 0.89 & 372 \\
Random Forest Optimal & Elastic Net Optimal  & \begin{tabular}[c]{@{}c@{}}0.82\\ (0.03)\end{tabular} & \begin{tabular}[c]{@{}c@{}}1.45\\ (1.15)\end{tabular} & 0.84 & 372 \\
Random Forest Optimal & Linear Model Optimal & \begin{tabular}[c]{@{}c@{}}0.8\\ (0.03)\end{tabular}  & \begin{tabular}[c]{@{}c@{}}1.75\\ (1.2)\end{tabular}  & 0.83 & 372 \\
Elastic Net Optimal   & Mkt                  & \begin{tabular}[c]{@{}c@{}}0.64\\ (0.04)\end{tabular} & \begin{tabular}[c]{@{}c@{}}3.16\\ (1.64)\end{tabular} & 0.75 & 372 \\
Elastic Net Optimal   & Base                 & \begin{tabular}[c]{@{}c@{}}1.08\\ (0.05)\end{tabular} & \begin{tabular}[c]{@{}c@{}}1.71\\ (1.54)\end{tabular} & 0.76 & 372 \\
Elastic Net Optimal   & Linear Model Optimal & \begin{tabular}[c]{@{}c@{}}0.98\\ (0.01)\end{tabular} & \begin{tabular}[c]{@{}c@{}}0.35\\ (0.3)\end{tabular}  & 0.99 & 372 \\
Linear Model Optimal  & Mkt                  & \begin{tabular}[c]{@{}c@{}}0.65\\ (0.04)\end{tabular} & \begin{tabular}[c]{@{}c@{}}2.82\\ (1.63)\end{tabular} & 0.76 & 372 \\
Linear Model Optimal  & Base                 & \begin{tabular}[c]{@{}c@{}}1.09\\ (0.05)\end{tabular} & \begin{tabular}[c]{@{}c@{}}1.4\\ (1.59)\end{tabular}  & 0.76 & 372 \\
Base                  & Mkt                  & \begin{tabular}[c]{@{}c@{}}0.49\\ (0.03)\end{tabular} & \begin{tabular}[c]{@{}c@{}}2.65\\ (1.5)\end{tabular}  & 0.67 & 372

    \\ \hline

\iffalse
& &  $\bar{\gamma} = 6$ & &  \\ \hline
$f_a$                                      & $f_b$                                      & Beta ($\beta$) & Alpha ($\alpha$) & $R^2$ & $N_{obs}$ \\ \hline
Mkt\textsuperscript{RF}   & Mkt                                        &    \begin{tabular}[c]{@{}c@{}}0.69\\ (0.05)\end{tabular}     &    \begin{tabular}[c]{@{}c@{}}3.85\\ (1.32)\end{tabular}      &   0.53     & 706       \\
Mkt\textsuperscript{RF}    & Mkt\textsuperscript{Base} &    \begin{tabular}[c]{@{}c@{}}1.00\\ (0.04)\end{tabular}     &    \begin{tabular}[c]{@{}c@{}}2.47\\ (.96)\end{tabular}      &   0.74   & 706       \\
Mkt\textsuperscript{RF}   & Mkt\textsuperscript{LM}   &    \begin{tabular}[c]{@{}c@{}}0.78\\ (0.04)\end{tabular}      &    \begin{tabular}[c]{@{}c@{}}3.78\\ (1.56)\end{tabular}       &    0.37   & 706       \\
Mkt\textsuperscript{LM} & Mkt                                        &   \multicolumn{1}{c}{\begin{tabular}[c]{@{}c@{}}0.43\\ (0.04)\end{tabular}}      &  \begin{tabular}[c]{@{}c@{}}3.06\\ (1.20)\end{tabular}        &  0.33     & 706       \\
Mkt\textsuperscript{Base}   & Mkt                                        &   \begin{tabular}[c]{@{}c@{}}0.65\\ (0.03)\end{tabular}     &    \begin{tabular}[c]{@{}c@{}}1.64\\ (0.96)\end{tabular}       &     0.64   & 706      \\ \hline

\fi

\end{tabular}
\end{table}

Table \ref{tab:4} reports results from running regressions of the machine learning reward-risk timing strategies on the market index and the other strategies. The intercepts (Jensen's $\alpha$’s) (Jensen, 1968) are positive and statistically significant in all cases, except for the base. The machine learning strategy has an annualized alpha of 3.37\% and a beta of only 0.57. The machine learning strategy over the base, linear model, and Elastic Net reward-risk timing has annualized alphas of 1.44\%, 1.75\%, and 1.45\%, respectively. For the comparisons, the alphas earned from using Elastic Net, linear model, and unconditional mean and recent return variance to forecast the excess return and volatility are smaller at 3.16\% 2.82\%, and 2.65\%, respectively.   

We also conduct formal tests of marketing timing including the (HM) (Henriksson and Merton, 1981) and Tre-Mauzy (TM) (Treynor and Mauzy, 1966) tests. The HM test adds a second term to the model, the up-market excess return. It measures the alpha that cannot be replicated by a mix of options and the market index.
\begin{equation}
    f_{t+1}^a = \alpha + \beta f_{t+1}^b + \gamma \max(0,f_{t+1}^b) + \epsilon_{t+1} ,
\end{equation}
where $\gamma$ measures the degree of market-timing ability. In the case strategy $b$ is the market index, a positive $\gamma$ would demonstrate market timing ability. The TM test has the additional squared excess market return term, for which the coefficient reflects the convexity achieved by exposure to the market.
\begin{equation}
    f_{t+1}^a = \alpha + \beta f_{t+1}^b + \gamma (f_{t+1}^b)^2 + \epsilon_{t+1} ,
\end{equation}

In Table \ref{tab:5} are the results from running the above regressions for the various strategies. The Random Forest optimal strategy has statistically significant coefficients for both the TM and HM tests. Elastic Net optimal and return-only strategies have statistically significant coefficients for the TM test but not the HM test. Unsurprisingly, the linear model and base strategies do not have large positive coefficients.

\begin{table}[!htbp]
\captionsetup{labelfont=bf,font=normalsize}
\caption{\textbf{Tests of Out-of-Sample Timing Ability}}

\centering
\label{tab:5}
\justifying{\small{\noindent In this table, we perform marketing timing statistical tests on various strategies. The HM test: $f_{t+1}^a = \alpha + \beta f_{t+1}^b + \gamma \max(0,f_{t+1}^b) + \epsilon_{t+1} $ and the TM test: $f_{t+1}^a = \alpha + \beta f_{t+1}^b + \gamma (f_{t+1}^b)^2  + \epsilon_{t+1}$ are run and the gamma coefficients are given. Heteroskedasticity-corrected $t$-statistics are in parentheses. The data are monthly and the sample period is 1989 to 2019.}}

\medskip
\centering
\begin{tabular}{lcc}

\\ \hline                                
%& &  $\bar{\gamma} = 4$ & &  \\ \hline
Strategy                                      & TM                                   &  HM \\ \hline
Base                     & \begin{tabular}[c]{@{}c@{}}0.004\\ (1.17)\end{tabular} & \begin{tabular}[c]{@{}c@{}}0.031\\ (0.36)\end{tabular} \\
\\
Linear Model Optimal     & \begin{tabular}[c]{@{}c@{}}0.006\\ (1.53)\end{tabular} & \begin{tabular}[c]{@{}c@{}}0.096\\ (1.06)\end{tabular} \\
Linear Model Returns     & \begin{tabular}[c]{@{}c@{}}0.007\\ (1.74)\end{tabular} & \begin{tabular}[c]{@{}c@{}}0.101\\ (1.08)\end{tabular} \\
Linear Model Volatility  & \begin{tabular}[c]{@{}c@{}}0.005\\ (1.36)\end{tabular} & \begin{tabular}[c]{@{}c@{}}0.056\\ (0.71)\end{tabular} \\
\\
Elastic Net Optimal      & \begin{tabular}[c]{@{}c@{}}0.008\\ (1.86)\end{tabular} & \begin{tabular}[c]{@{}c@{}}0.114\\ (1.24)\end{tabular} \\
Elastic Net Returns      & \begin{tabular}[c]{@{}c@{}}0.008\\ (2.02)\end{tabular} & \begin{tabular}[c]{@{}c@{}}0.115\\ (1.24)\end{tabular} \\
Elastic Net Volatility   & \begin{tabular}[c]{@{}c@{}}0.005\\ (1.37)\end{tabular} & \begin{tabular}[c]{@{}c@{}}0.056\\ (0.71)\end{tabular} \\
\\
Random Forest Optimal    & \begin{tabular}[c]{@{}c@{}}0.008\\ (2.33)\end{tabular} & \begin{tabular}[c]{@{}c@{}}0.134\\ (1.67)\end{tabular} \\
Random Forest Returns    & \begin{tabular}[c]{@{}c@{}}0.009\\ (2.25)\end{tabular} & \begin{tabular}[c]{@{}c@{}}0.128\\ (1.49)\end{tabular} \\
Random Forest Volatility & \begin{tabular}[c]{@{}c@{}}0.004\\ (1.27)\end{tabular} & \begin{tabular}[c]{@{}c@{}}0.041\\ (0.52)\end{tabular}

    \\ \hline

\end{tabular}
\end{table}

The next finding is that our strategies survive transaction costs, given in Table \ref{tab:tc}. Specifically, we evaluate our portfolio allocation strategy for the reward-risk timing portfolios when accounting for empirically realistic transaction costs as in (Moreira and Muir, 2017). Strategies that capture reward-risk timing but reduce trading activity include capping the strategy’s leverage at 1 compared to the case with a weight limit of 1.5. These leverage limits reduce trading and hence total transaction costs. We report the average absolute change in monthly weights, expected return, and Jensen's alpha of each strategy before transaction costs. The next columns contain the alphas when including various transaction cost assumptions. Finally, the last column derives the implied trading costs in basis points such that the alphas are zero in each of the cases.

%\end{multicols}
\begin{table}%[h] 
\captionsetup{labelfont=bf,font=normalsize}
\caption{\textbf{Transaction Costs of Machine Learning Portfolio Allocation}}

\justifying{\small{\noindent In this table, we evaluate our reward-risk timing strategies for the market when including transaction costs. Lower leverage limits reduce trading activity. Specifically, we consider restricting risk exposure to be between 0 and 1 (i.e., no leverage) or 1.5. The alphas are reported with these assumptions. Following Moreira and Muir (2017), the 1bp cost comes from Fleming et al.\ (2003), the 10bps is from Frazzini, Israel, and Moskowitz (2015) when trading approximately 1\% of daily volume, and the next column adds an additional 4bps to cover for transaction costs increasing in high-volatility episodes. The last column backs out the implied trading costs in basis points needed to drive the alphas to zero in each of the cases.}}

\medskip

\centering
\label{tab:tc}
%\resizebox{\textwidth}{!}{
\begin{tabular}{llllllllc}
\hline
                                                            &                              &                &        &           \multicolumn{4}{l}{$\alpha$ After Trading Costs} \\ 
Weight                                                                  & $| \Delta w |$ & $E[R]$ & $\alpha$ & 1bps     & 10bps     & 14bps     & Break Even    \\ \hline
 \multicolumn{8}{c}{ $\bar{\gamma} = 4$}  \\ \hline
Random Forest Optimal 1.5 & 0.21 & 13.42 & 3.37 & 3.34 & 3.11 & 3.01 & 132.16 \\
Elastic Net Optimal 1.5   & 0.28 & 14.01 & 3.16 & 3.12 & 2.82 & 2.68 & 93.46  \\
Linear Model Optimal 1.5  & 0.28 & 13.84 & 2.82 & 2.78 & 2.47 & 2.33 & 81.70   \\
Base 1.5                  & 0.29 & 11.63 & 2.65 & 2.62 & 2.31 & 2.17 & 77.94  \\
Random Forest Optimal 1   & 0.11 & 10.75 & 1.96 & 1.95 & 1.83 & 1.78 & 148.80  \\
Elastic Net Optimal 1     & 0.14 & 11.01 & 2.00    & 1.98 & 1.83 & 1.77 & 120.6  \\
Linear Model Optimal 1    & 0.14 & 10.99 & 1.88 & 1.86 & 1.70  & 1.63 & 106.94 \\
Base 1                    & 0.15 & 9.63  & 1.68 & 1.66 & 1.51 & 1.44 & 95.94  \\
\hline

 \multicolumn{8}{c}{ $\bar{\gamma} = 6$}  \\ \hline
Random Forest Optimal 1.5 & 0.24 & 11.00   & 2.83 & 2.8  & 2.54 & 2.42 & 97.06  \\
Elastic Net Optimal 1.5   & 0.33 & 12.55 & 3.42 & 3.38 & 3.01 & 2.85 & 84.28  \\
Linear Model Optimal 1.5  & 0.34 & 12.50  & 3.22 & 3.18 & 2.81 & 2.64 & 77.60   \\
Base 1.5                  & 0.33 & 10.22 & 2.68 & 2.64 & 2.28 & 2.12 & 67.90   \\
Random Forest Optimal 1   & 0.14 & 9.90   & 2.25 & 2.23 & 2.08 & 2.01 & 131.16 \\
Elastic Net Optimal 1     & 0.18 & 10.29 & 2.10  & 2.08 & 1.88 & 1.79 & 92.46  \\
Linear Model Optimal 1    & 0.19 & 10.18 & 1.88 & 1.86 & 1.65 & 1.56 & 80.71   \\
Base 1                    & 0.19 & 8.70   & 1.77 & 1.74 & 1.54 & 1.45 & 75.34 \\
\hline    

\end{tabular}
%}
\end{table}
%\begin{multicols}{2}

The results indicate that machine learning reward-risk timing survives transactions costs, even with high volatility episodes where such fees rise. Overall, the annualized alpha of the reward-risk timing portfolio allocation strategy decreases slightly, but is still very large. Reward-risk timing with machine learning does not require extreme leverage or drastic portfolio rebalancing to be profitable.

The empirical results overall indicate a significant advantage in using machine learning for portfolio allocation. With only standard predictor variables, reward-risk timing with machine learning models offers economically substantial improvements in risk-adjusted returns (28\% increase in Sharpe ratio). Statistically significant positive alphas of 3.4\% are found as a result of the superior forecasting ability of machine learning. Finally, realistic trading costs are applied to gain further insight on real-life applicability, showing alphas remain large. %With this evidence in mind, it is also valuable to look from a theoretical perspective at why the strategy outperforms.

\section{Conclusion} \label{sec:conclusion}

Machine learning portfolio allocation offers large risk-adjusted returns and utility gains and is feasible to implement in real-time. We perform both return- and volatility-timing, or reward-risk timing, with and without machine learning, showcasing the relative advantage the machine learning models Random Forest and Elastic Net can provide. Furthermore, our strategy's performance is informative about the alpha generation process for actively managed portfolios.

At the same time, there are possibilities for improvements. Other machine learning methods like deep neural networks may allow trading some interpretability for performance gains. Using predictors beyond lagged payout yields and risk-free rates may also be beneficial. Additionally, this strategy on daily or weekly data may have the benefit of catching sharp drops in the market. Since one of our goals here was to show that machine learning has an advantage in finance and portfolio allocation outside the context of big data, the results with standard variables are promising.

\bibliographystyle{unsrt}  


\begin{thebibliography}{1}



\bibitem{17.5}
Henrique BM., Sobreiro VA, Kimura H. 2019.
\newblock Literature review: Machine learning techniques applied to financial
market prediction.  
\newblock {\em Expert Systems With Applications} 124:226-51. 

\bibitem{15}
Gu S, Kelly BT, Xiu D. 2020. 
\newblock Empirical Asset Pricing via Machine Learning. 
\newblock {\em The Review of Financial Studies} 33(5):2223–73.

\bibitem{14}
Goyal A, Welch I. 2008.
\newblock A Comprehensive Look at The Empirical Performance of Equity Premium Prediction.
\newblock {\em The Review of Financial Studies} 21(4):1455-508. 

\bibitem{4}
Boudoukh J, Michaely R, Richardson M. 2007. 
\newblock On the Importance of Measuring Payout Yield: Implications for Empirical Asset Pricing.
\newblock {\em The Journal of Finance} 62:877-915.

\bibitem{44}
Grinold RC, Kahn RN. 1999. 
\newblock Active Portfolio Management: A Quantitative Approach for Producing Superior Returns and Selecting Superior Returns and Controlling Risk. 
\newblock McGraw-Hill Library of Investment and Finance.

\bibitem{29}
Merton R. 1981.
\newblock On Market Timing and Investment Performance. I. An Equilibrium Theory of Value for Market Forecasts.
\newblock {\em The Journal of Business}  54(3):363-406.

\bibitem{8}
Campbell JY, Thompson SB. 2008. 
\newblock Predicting Excess Stock Returns Out of Sample: Can Anything Beat the Historical Average?.
\newblock {\em The Review of Financial Studies} 21(4):1509-31.

\bibitem{30}
Moreira A, Muir T. 2017.
\newblock Volatility‐Managed Portfolios.
\newblock {\em The Journal of Finance} 69(2):1611-44.

\bibitem{22}
Kandel S, Stambaugh RF. 1996.
\newblock On the Predictability of Stock Returns: An Asset-Allocation Perspective.
\newblock {\em The Journal of Finance} 51(2):385-424.

\bibitem{20}
Johannes M, Korteweg A, Polson N. 2014.
\newblock Sequential Learning, Predictability, and Optimal Portfolio Returns.
\newblock {\em The Journal of Finance} 69(2):611-644. 

\bibitem{12}
Fleming J, Kirby C, Ostdiek B. 2001. 
\newblock The Economic Value of Volatility Timing.
\newblock {\em The Journal of Finance} 56:329-52.

\bibitem{31}
Moreira A, Muir T. 2019.
\newblock Should Long-Term Investors Time Volatility?.
\newblock {\em The Journal of Financial Economics} 131(3):507-27.

\bibitem{25}
Liu F, Tang X, Zhou G. 2019.
\newblock Volatility-Managed Portfolio: Does It Really Work?,
\newblock {\em The Journal of Portfolio Management} 46(1):38-51. 

\bibitem{2511}
Marquering W, Verbeek M. 2004.
\newblock The Economic Value of Predicting Stock Index Returns and Volatility,
\newblock {\em Journal of Financial and Quantitative Analysis} 39(2):407-29. 

\bibitem{23}
Kirby C, Ostdiek B. 2012.
\newblock It’s All in the Timing: Simple Active Portfolio Strategies that Outperform Naive Diversification.
\newblock {\em Journal of Financial and Quantitative Analysis} 47(2):437–67.

\bibitem{16}
Nystrup P, Hansen BW, Madsen H, Lindström E. 2016.
\newblock Detecting change points in VIX and S\&P 500: A new approach to dynamic asset allocation.
\newblock {\em J Asset Manag} 17:361-74.

\bibitem{1}
Bailey D, Borwein J, De Prado MZ, Zhu QJ. 2017.
\newblock The probability of backtest overfitting. 
\newblock {\em Journal of Computational Finance} 20:39–69.

\bibitem{34}
Zou H, Hastie T. 2005. 
\newblock Regularization and variable selection via the elastic net. 
\newblock {\em J. R. Statist. Soc. B} 67:301–20.

\bibitem{6}
Breiman L. 2001. 
\newblock Random Forests.
\newblock {\em Machine learning} 45:5–32.

\bibitem{17}
Hastie T, Friedman J, Tibshirani R. 2017.
\newblock The Elements of Statistical Learning 2nd ed.,
\newblock Springer.

\bibitem{24}
Liaw A, Wiener M. 2002.
\newblock Classification and Regression by randomForest.
\newblock {\em R News} 2:18-22. 

\bibitem{21}
Johannes M, Polson N, Stroud J. 2004.
\newblock Sequential optimal portfolio performance: Market and volatility.
\newblock Working Paper, Columbia University, University of Pennsylvania, and University of Chicago. 

\bibitem{11}
Fama E, French KR. 1988b. 
\newblock Dividend yields and expected stock returns.
\newblock {\em Journal of Financial Economics} 222:3-25.

\bibitem{10}
Engle R. 1982.
\newblock Autoregressive conditional heteroskedasticity with estimates of the variance of U.K. inflation.
\newblock {\em Econometrica} 50:987-1008.

\bibitem{5}
Bollerslev T, 1986.
\newblock Generalized Autoregressive Conditional Heteroskedasticity.
\newblock {\em Journal of Econometrics}  31:307-27.

\bibitem{27}
Matteson DS. and Ruppert D. 2015.
\newblock Statistics and Data Analysis for Financial Engineering, 2nd Ed.

\bibitem{4.5}
Boudoukh J, Richardson M, Whitelaw RF. 2006.
\newblock The Myth of Long-Horizon Predictability.
\newblock {\em Review of Financial Studies} 21(4):1577-605.

\bibitem{4.7}
Shapley LS. 1953.
\newblock A value for n-person games.
\newblock {\em Contributions to the Theory of Games} 28(2):307–17.

\bibitem{4.6}
Lundberg SM, Lee SI. 2017.
\newblock A Unified Approach to Interpreting Model
Predictions.
\newblock {\em Advances in Neural Information Processing Systems} 21(4):4765–74.


\bibitem{19}
Jensen MC. 1968. 
\newblock The Performance of Mutual Funds in the Period 1945-1964.
\newblock {\em The Journal of Finance} 23(2):389-416.

\bibitem{18}
Henriksson RD, Merton R. 1981. 
\newblock On Market Timing and Investment Performance. II. Statistical Procedures for Evaluating Forecasting Skills,
\newblock In {\em The Journal of Business} 54 No. 4, 513-33.

\bibitem{33.5}
Treynor JL, Mauzy K. 1966. 
\newblock Can Mutual Funds Outguess the Market?. 
\newblock {\em Harvard Business Review} 44:347-68.

\bibitem{34}
White H. 1980. 
\newblock A heteroskedasticity-consistent covariance matrix estimator and a direct test for heteroskedasticity. 
\newblock {\em Econometrica} 48:817-38.

\bibitem{3433}
Fleming J, Kirby C, Ostdiek B, 2003. 
\newblock The economic value of volatility timing
using “realized” volatility. 
\newblock {\em Journal of Financial Economics} 67:473–509.

\bibitem{13}
Frazzini A, Israel R, Moskowitz T. 2015.
\newblock Trading costs of asset pricing anomalies.
\newblock Working paper, AQR Capital Management.

\bibitem{33}
Samuelson PA. 1969. 
\newblock Lifetime Portfolio Selection by Dynamic Stochastic Programming. 
\newblock {\em Review of Economics and Statistics} 51:239-46.

\bibitem{.56}
Brandt M. 1999.
\newblock Estimating Portfolio and Consumption Choice: A Conditional Euler Equations Approach.
\newblock {\em The Journal of Finance} 54(5):1609–45.

\bibitem{14.5}
Gron A, Jørgensen BN, Polson NG. 2011.
\newblock Optimal portfolio choice and stochastic volatility.
\newblock {\em Applied Stochastic Models in Business and Industry} 28(1):1-15. 

\bibitem{7}
Breiman L, Friedman JH, Olshen RA, Stone CJ. 1984. 
\newblock Classification and regression trees.
\newblock CRC press.

\bibitem{32}
Murphy K. 2012. 
\newblock Machine Learning - a Probabilistic Perspective.
\newblock MIT Press.














%\bibitem{.5}
%Ait-Sahalia, Y., and Michael Brandt, 2001.
%\newblock Variable Selection for Portfolio Choice.
%\newblock {\em The Journal of Finance} 56(4):1297–351.





%\bibitem{2}
%Bao, Yong, 2009,
%\newblock Estimation Risk-Adjusted Sharpe Ratio and Fund Performance Ranking Under a General Return Distribution,
%\newblock In {\em Journal of Financial Econometrics} 7 No. 2, 152-173.

%\bibitem{3}
%Barroso, Pedro, Pedro Santa-Clara. 2014.
%\newblock Momentum has its moments.
%\newblock {\em Journal of Financial Economics} 116:111-20.













%\bibitem{9}
%Cederburg, Scott, Michael S. O’Doherty, Feifei Wang, Xuemin (Sterling) Yan, 2019, 
%\newblock On the performance of volatility-managed portfolios, 
%\newblock Forthcoming  In {\em Journal of Financial Economics}.










%\bibitem{16}
%Hallac, David, Peter Nystrup, and Stephen Boyd. 2018.
%\newblock Greedy Gaussian segmentation of multivariate time series.
%\newblock {\em Advances in Data Analysis and Classification} 13(3):727–51.


















%\bibitem{26}
%Markowitz, Harry, 1952, 
%\newblock Portfolio Selection
%{\em The Journal of Finance}, Vol. 7, No. 1., 77-91.



%\bibitem{28}
%Merton, Robert. 1969.
%\newblock Lifetime Portfolio Selection under Uncertainty: The Continuous-Time Case.
%\newblock {\em The Review of Economics and Statistics} 51(3):247-57.

%\bibitem{18}
%Sugiura, Nariaki. 2007. 
%\newblock Further analysts of the data by akaike' s information criterion and the finite corrections.
%\newblock {\em Commun. Stat.} 7:13–26.









%\bibitem{35}
%Yoo, Paul D., Maria H. Kim, and Tony Jan. 2005.
%\newblock Machine Learning Techniques and Use of Event Information for Stock
%Market Prediction: A Survey and Evaluation.  
%\newblock International Conference on Computational Intelligence for Modeling, Control %and Automation. 


\end{thebibliography}

\newpage

\appendix

\noindent {\LARGE \textbf{Appendix}}

\section{Optimal weights}

Samuelson (1969) showed the optimal investment fraction in the risky asset to maximize the expected utility of wealth is given by:
\begin{equation} \label{eq:3}
{ \large w^*_t = \frac{\mu-r^f_t}{\gamma \sigma^2}}
 \normalsize.
\end{equation}
It is well known that the investment opportunities are not constant throughout time. Therefore, consider the following model where the market expected return and volatility change according to two non-linear functions of lagged predictor variables and volatilities. 
\begin{gather}
r_t = \mu_t + \sigma_t \cdot z_t \\
\mu_t = g_t \left ( \vec{x}_{t-1} \right ) + \epsilon_t \\
\log(\sigma_t^2) = h_t \left ( \vec{v}_{t-1}  \right ) + s_t,
\end{gather}
\noindent where $\vec{x}_{t-1}$ is the vector of predictor variables for the excess return model, $\vec{v}_{t-1}$ is the vector for the volatility model, $z_t$, $\epsilon_t$, and $s_t$ are potentially correlated normal random variables with mean zero, $E[z_t|z_{t-1}] = E[z_t]$, $E[\epsilon_t|\epsilon_{t-1}] = E[\epsilon_t]$, and $E[s_t|s_{t-1}] = E[s_t]$. Functions $g_t$ and $h_t$ are unknown and to be estimated. In certain stylized cases, there exist closed-form solutions to multi-period investment problems when variables at the current time are unknown. As Johannes et al.\ (2004) point out, however, for an analytical solution, expected returns can be unknown only if the current volatility is known, for instance, by the quadratic variation process. Because both future returns and volatility are predicted, to solve the optimal portfolio problem, we follow the existing literature and simplify the allocation problem by considering a single-period problem:
\begin{equation}
    J(\mathcal{F}_{t-1}) = \max_{w_t}E[U(W_{t})|\mathcal{F}_{t-1}] \\ 
    = \max_{w_t}\int U(W_t)P(r_t|\mathcal{F}_{t-1})dr_t, 
\end{equation}
where $P(r_t | \mathcal{F}_{t-1})$ is the predictive distribution of future returns and $\mathcal{F}_{t-1}$ is the information set known at time $t-1$. This is similar to the approach taken in Kandel and Stambaugh (1996) and Johannes et al.\ (2014). 

The difference between single and multi-period problems is that in the latter, hedging demands arise from changes in variables determining the attractiveness of future investment opportunities. Brandt (1999) showed that hedging demands are typically very small terms in the optimal weight. Additionally, portfolio choice will be myopic if the investor has power utility and returns are IID.

To derive the optimal portfolio weight, let us assume that $U(\cdot)$ is twice differentiable, monotonically increasing, and concave (which is the case for the power utility investor). Then by Eq. 3, the optimal portfolio is given by the first order condition
\begin{equation} \label{eq:9}
    E[U^{'} (W_{t})(R_t-R_t^f)|\mathcal{F}_{t-1}] = 0,
\end{equation}
\noindent where $R_t$ denotes $\exp(r_t) - 1$, $R_t^f$ is $\exp(r_t^f) - 1$, and the expectation is taken over the predictive distribution of future returns. By the definition of covariance and Eq. \ref{eq:9}, 
\begin{equation} \label{eq:10}
    cov[U^{'}(W_t),R_t-R_t^f|\mathcal{F}_{t-1}] + E[U'(W_t)|\mathcal{F}_{t-1}]E[R_t-R_t^f|\mathcal{F}_{t-1}] = 0,
\end{equation}
To separate the effects of risk and return on utility, realize that $R_t$ has a stochastic volatility mixture distribution (Gron et al., 2011). In this case, a generalization of Stein's lemma (see Appendix B) allows us to re-write the covariance term as 
%To separate the effects of risk and return on utility, realize that the predictive distribution of $r_t-r_t^f$ is a mixture distribution. In this case, a generalization of Stein's lemma (see Johannes et al.\ (2004)) allows us to re-write the covariance term as 
\begin{gather}
    cov[U^{'} (W_t),R_t-R_t^f|\mathcal{F}_{t-1}] = E^Q[U^{''}(W_t)|\mathcal{F}_{t-1}] cov[W_{t},R_t|\mathcal{F}_{t-1}] \nonumber \\ = 
    w_t E^Q[U^{''}(W_t)|\mathcal{F}_{t-1}]var[R_t|\mathcal{F}_{t-1}],
\end{gather}
where $Q$ represents the size-biased volatility-adjusted distribution. Solving for the optimal weight,
\begin{equation} \label{eq:12}
    w_t^* = \frac{E[R_t-R_t^f|\mathcal{F}_{t-1}]}{\bar{\gamma} \cdot var[R_t |\mathcal{F}_{t-1}] },
\end{equation}
where $\bar{\gamma} = -E[U^{'}(W_t)|\mathcal{F}_{t-1}]/E^{Q}[U^{''}(W_t)|\mathcal{F}_{t-1}]$. This provides a justification for using a conditional mean-variance rule.

As a final case, consider constant-mean returns and time-varying volatility:
\begin{gather}
r_t = \mu + \sigma_t \cdot z_t \\
\log(\sigma_t^2) = h_t \left ( \vec{v}_{t-1}  \right ) + s_t
\end{gather}
Starting from Eq. \ref{eq:10}, using the fact that $E[R_t-R_t^f|\mathcal{F}_{t-1}] = E[R_t-R_t^f]$, and applying the same logic, the optimal weight is given by
\begin{equation} \label{eq:15}
    w_t^* = \frac{E[R_t-R_t^f]}{ \bar{\gamma} \cdot var[R_t |\mathcal{F}_{t-1}] }.
\end{equation}

The two functions $g_t(\mathcal{F}_{t-1}) = R_t-R_t^f$ and $h_t(\mathcal{F}_{t-1}) = \log(\sigma^2_{t})$ give the expected excess return and variance, respectively, at time $t$ given the information set $\mathcal{F}_{t-1}$ at the previous time. In this paper, we learn $g_t$ and $h_t$ with the machine learning algorithm Random Forest discussed in Section \ref{Random Forest}.

\section{Stein’s lemma for stochastic volatility}
\numberwithin{equation}{section}

Let $X$ be a random variable with a stochastic volatility so that $X|\sigma$ is distributed $N(\mu, V^2 \sigma  )$ and $\sigma$ has density $p(\sigma)$ that is non-negative only for $\sigma \geq 0$. Let $g(X)$ be the differentiable function of $X$ such that $E[|g(X)|]<\infty$. Suppose that $ 0< E[\sigma] < \infty $. If $ ( X, Y | \sigma ) $ are bivariate Normal random variables then 
\begin{equation}
    cov[g(X),Y]= E^Q[g'(X)] cov[X,Y],
\end{equation}
\noindent where $E^Q$ is the expectation taken under the measure induced by size-biasing $q(\sigma)= \sigma p(\sigma)/ E[ \sigma ]$. For a proof see Gron et al.\ (2011).

\section{Decision tree algorithms}

%\numberwithin{algorithm}{section}

Algorithm C1 details how to build a regression tree in a Random Forest and is a greedy algorithm (Breiman et al., 1984). We refer to the recursive version in (Murphy, 2012).
\bigskip

%\newcounter{algorithm}
\setcounter{algocf}{0}
\renewcommand{\thealgocf}{C\arabic{algocf}}

\begin{algorithm}[H] 
\label{alg:3}
%\begin{varwidth}[t]{0.4\textwidth}
%\SetAlgoLined
Initialize stump node, $N_1(0)$. $N_k(d)$ is the $k$th node at depth $d$. $S$ denotes the data, and $C$ is the set of unique labels. \\
\bigskip
function fitTree($N_k(d)$, $S$, $d$)
  \begin{enumerate}
  \item The prediction of the $N_k(d)$ node is the average value of its observations, $ \frac{1}{|N_k(d)|} \sum_{i\in N_k(d)} y_i $ 
  
  \item Define the cost function as the sum of squared differences from the mean: $cost(\{x_{i},y_{i}\}) = \sum_{i \in \{x_i,y_i\} } (y_i - \bar{y})^2 $, where $\bar{y} = \frac{1}{|\{x_i,y_i\}|} \sum_{i \in \{x_i,y_i\} } y_i $ \\ is the mean of the response variable in the specified set of data.
  
  \item Select the optimal split: \\ $(j^*, t^*) =  \argmin_{j \in \{1,..,m\} } \min_{t \in \mathcal{T}_j } (cost(\{ x_{i}, y_{i} :x_{ij} \leq t  \}) + cost(\{ x_{i}, y_{i} :x_{ij} > t \}) )$.
  
  $S_{left} = \{ x_{i}, y_{i} :x_{ij} \leq t  \}$, $S_{right} = \{ x_{i}, y_{i} :x_{ij} > t \}$.
  
  \item{\SetAlgoNoLine
     \eIf{notworthSplitting($d$, $cost$, $S_{left}$,$S_{right}$)}{
         \Indp  \ \ \ \ \  return $N_k(d)$
     }{
        \Indp{\Indp{
        Update the nodes: \\
         $N_1(d+1) = $ fitTree($N_k(d)$, $S_{left}$, $d+1$) \\
         $N_2(d+1) = $ fitTree($N_k(d)$, $S_{right}$, $d+1$) \\
         \ \ \ \ \   return $N_k(d)$ 
        }}
     }
  }      
 \end{enumerate}
 
\KwResult{The regression tree model $f(\vec{x}) = \sum_{m=1}^{D} w_m \mathds{1}\{ \vec{x} \in S_m \} $, where $w_m = \frac{1}{|S_m|} \sum_{i\in S_m} y_i  $ and $D$ is the number of regions }
 \caption{Regression Tree}
% \end{varwidth}
\end{algorithm}
\medskip
\noindent The function $notworthSplitting$($d$, $cost$, $S_{left}$,$S_{right}$) contains stopping heuristics to prevent overfitting. In our case, the function value is true if the fraction of examples in either $S_{left}$ or $S_{right}$ is less than $s_{min}$, the minimum fraction of observations in a node for a split determined by the user's parameter optimization, or if the number of terminal nodes $D$ is equal to $k_{max}$, the maximum number of terminal nodes. An important note is that the $S_{min}$ threshold is applied to the current node. For instance, a node can contain 5 observations out of 100 in the data even if $S_{min} = 0.9$, but any further splits from that node will not be made since $5/100 < 0.9$.

For the reward Random Forest model, which estimates the excess return, the values we set for $s_{min}$, $k_{max}$, the number of trees, and the number of variables to select from at each split ($m$) are 0.95, 2, 500, and 4, respectively. For the volatility Random Forest model, the respective values are 0.01, 12, 500, and 4.

\end{document}